\documentstyle[12pt,twoside,fleqn,espcrc1,epsf]{article}

\newcommand{\beq}{\begin{equation}}
\newcommand{\eeq}[1]{\label{#1} \end{equation}}
\newcommand{\beqar}{\begin{eqnarray}}
\newcommand{\eeqar}[1]{\label{#1} \end{eqnarray}}

\begin{document}
\begin{flushright}
CU-TP-878
\end{flushright}
{\Large Baryon Junction Stopping at the SPS and RHIC via HIJING/B}\\[2ex]

S.E. Vance$^a$\footnote{Presented at QM97, Tsukuba, Japan}, M.Gyulassy$^a$ 
  X.N. Wang$^b$\\[2ex]
{$^a$ Physics Department, Columbia
    University, New York, NY, 10027}\\
{$^b$Nuclear Science Division, LBNL, Berkeley, CA 94720}\\[2ex]



\begin{abstract}
Baryon stopping at the SPS and RHIC energies is calculated
by introducing a new baryon junction mechanism into HIJING.
The exchange of a baryon junction, according to Regge
phenomenology,  leads to a $\cosh[(y-y_{CM})/2]$ rapidity dependence and an
$1/\sqrt[4]{s}$ energy dependence of the inclusive baryon cross section. 
This  baryon junction dynamics also leads naturally to enhanced 
$p_T$ broadening in $pA$ and $AA$ together with
enhanced mid-rapidity hyperon production.
\end{abstract}

\section{Introduction}
The systematic study of baryon stopping and hyperon
production in $pp, pA,$ and $AA$ collisions
is essential in order to differentiate the sought after new 
physics of dense equilibrated matter from non-equilibrium
multi-particle production dynamics\cite{gyu_qm95}. 
Baryon stopping 
refers to the transport of baryon number in rapidity space
away from the nuclear fragmentation 
regions and is measured through the single inclusive rapidity distribution
of protons and hyperons.  As  discussed in
\cite{topor_95,gyu_stp97}, recent SPS data shows a high degree of 
 baryon stopping and anomalous hyperon  production.  
These findings are based on comparisons of 
predictions using the  HIJING \cite{hijing} and VENUS \cite{venus} 
Monte Carlo  event generators with  SPS data for 
$p+A$ and $A+B$ collisions.
In particular, the
baryon stopping mechanism modeled in HIJING was found to underestimate 
the stopping in heavy nuclear collisions,
while
VENUS model was  able to reproduce the baryon stopping and strange baryon 
production by introducing
a new non-equilibrium  dynamical mechanism called the double string.
The mechanism in HIJING, adapted from 
 the LUND Fritiof \cite{fritiof} and dual parton 
model (DPM) \cite{dpm_94}, involves diquark-quark 
string excitations followed by string breaking.
The final valence baryons  in this model
always emerge within about a unit of rapidity from
those of the unbroken diquarks.
This association of the baryon with
the end point diquark creates a characteristic baryon rapidity 
distribution where the baryons are concentrated near  the beam and 
target fragmentation regions as shown in Fig 1a.  
The extrapolation of the $pp$ physics of HIJING to $PbPb$ is seen
to underestimate the mid-rapidity valence baryon yield by a factor of two
in Fig 1b.  VENUS avoids this
``fly trap'' by postulating a specific model 
of diquark breakup, leading to double strings.
The double string emerging from a broken diquark 
is assumed to loop around a valence quark from the other
fragmentation region.
The produced baryon  is then associated with 
the fragmentation of the double string near its
 $180^{o}$ 
bend. The parameters of this mechanism are adjusted from $pA$
and suffice to reproduce the $AA$ data. We emphasize that
the analytic one parameter Multi-Chain-Model\cite{date} also
reproduces well\cite{gyu_stp97} 
the valence proton distributions based on fitting $pA$.
The advantage of a Monte Carlo event generator is the ability
to test via other observables such as hyperon production and $p_\perp$
systematics the consistency of specific dynamical assumptions
with a wide range of data.

Another model of baryon transport was proposed by
Kopeliovich and Zakharov \cite{kop_89} based upon pQCD considerations.
That mechanism 
has been recently included in a new version of the 
DPM \cite{dpm_dqb}.   In that  model, the leading diquark is broken by a
color exchange which changes its color state from a $\{\bar{3}\}$ to a 
$\{6\}$.  This allows one of the valence quarks to flow into the central
rapidity region and act as the seed for
the production of the final  baryon number.   The inclusive
cross section of this
mechanism is characterized by a  $\cosh{y/2}$ rapidity dependence and an 
$1/\sqrt[4]{s}$ energy dependence which follows from
the assumed  $1/\sqrt{x}$ distribution of the valence quarks in a proton.  
In practice this mechanism is  similar to that invoked in VENUS
in that the fragmentation of the now two independent quarks from the diquark
is similar to that from two independent strings.
As in VENUS this configuration allows for a
factor of 2 enhancement in the production of strange baryons. 
Adding a component of about $20\%$ for each nucleon-nucleon collision, 
with the
assumption that once the diquark is broken then it remains broken,
DPM is then also able to  reproduce the SPS data.

In this work we test a third model of baryon transport
recently proposed by Kharzeev\cite{khar_bj96}
and based on a long dormant 
Rossi-Veneziano baryon junction Regge exchange model\cite{rossi_77}. 
In this approach, the baryon number is traced by a non-perturbative
baryon junction of topological nature. The special theoretical appeal of that
idea arises from the form of the baryon wavefunction in QCD. 
The requirement of gauge invariance of 
the nonlocal operator creating a baryon naturally leads to the concept of a
baryon junction:
\beqar
\nonumber
B(x_1,x_2,x_3,x_J) = 
\epsilon^{ijk} \left [ P \exp \left (ig \int_{x_1}^{x_J} dx^{\mu} 
A_{\mu} \right ) q(x_1) \right ]_i \left [P \exp \left (ig \int_{x_2}^{x_J} 
dx^{\mu} A_{\mu} \right ) q(x_2) \right ]_j \\ \times
\left [P \exp \left (ig \int_{x_3}^{x_J} dx^{\mu} A_{\mu} \right ) q(x_3)
\right ]_k.
\eeqar{baryon}
The baryon junction  is a  vertex at $x_J$ where the three gluon        
Wilson lines emanating from the three valence quarks (in SU(3))
must join in order to form a gauge invariant operator.
In a highly excited baryonic state, the three valence quarks fragment
via multiple $q\bar{q}$ into mesons leaving
three sea quarks eventually around the junction to form
the observed final baryon. This is the sense in which  the junction
traces the baryon number. Being a purely gluonic configuration,
the junction may be easily transported into the mid-rapidity region.
The fragmentation of the end point valence quarks naturally leads  to 
three beam jets, very similar to the previous mechanisms considered. 
The primary advantage of this mechanism is that
Regge phenomenology can be used to estimate the
inclusive inelastic cross section for this process.
From $p\bar{p}$ and $pp$ data, the junction-anti-junction 
exchange is characterized by a trajectory with intercept,
$\alpha_{J\bar{J}}(0)\approx 1/2$. This leads immediately to  
a $\cosh{y/2}$ rapidity dependence and an $1/\sqrt[4]{s}$ 
energy dependence. 

The unique prediction of this mechanism
is that fragmentation of the valence quarks
down to the junction is expected to 
enhance hyperon production by a factor of 3 just from the random
combinatorics of $s$ vs $u,d$ pair production.
In addition, from the random addition
of three sea quarks, the  transverse momentum of the final baryon
is automatically enhanced by a factor of $\sqrt{3}$.  

Given these interesting coupled consequences of the baryon exchange
mechanism, we have chosen to implement this  model in
a new version 
of HIJING, called  HIJING/B\cite{sev_bj98}.

\begin{figure}[t]
\begin{center}
\leavevmode\epsfysize=4.0in
\epsfxsize=5.0in
\epsfbox{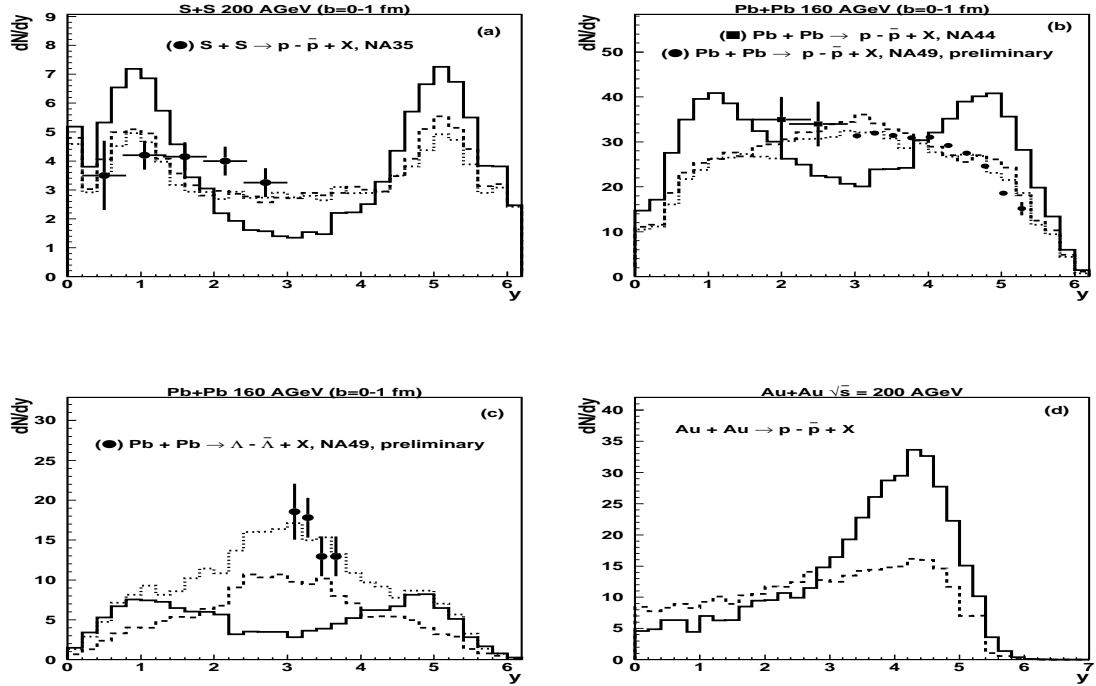}
\end{center}
\vspace{-0.75in}
\caption{A comparison of HIJING (solid), HIJING/B (dashed) and HIJING/B with
``ropes'' (dotted) with various experimental data.}
\end{figure}
%
%

\section{HIJING/B}
In HIJING/B, baryon junction stopping is implemented using a ``Y'' string 
configuration for the excited baryons
in which the baryon junction and produced baryon
are distributed as noted above. The resulting three
beam jets are treated as $q\bar{q}$ configurations.
For each nucleon-nucleon interaction, a $\sim 60\%$ 
probability is given that only one of the two nucleons 
will have its baryon junction stopped 
in the central rapidity region.  Baryon excitations without junction
exchange are taken to be standard $(qq)-q$ strings for fragmentation purposes.
Due to kinematical considerations, junction exchange
is allowed only if the invariant mass exceeds $m \ge 4.5$ GeV
so that the  three beam jets have room to decay. At SPS energies
this trivial kinematic constraint limits considerably the number of junction
exchanges allowed. 
The junction exchange probability is  fitted so as to reproduce
the $pp$ valence 
proton data. 
For multiple collisions, we assume that once the 
baryon junction is exchanged, it remains stopped in 
subsequent soft interactions.    

A comparison of the prediction of HIJING and HIJING/B with SPS
data \cite{na35,na44,na49} is given in Fig 1a,b,c.   
In part (a), (b)  we show the valence proton 
$(p-\bar{p})$ rapidity distribution for, $S+S$ and $Pb+Pb$, respectively. 
Whereas HIJING under-predicts the baryon stopping, HIJING/B is 
shown to provide a large amount of stopping.  
The valence hyperon $(\Lambda-\bar{\Lambda})$ 
rapidity distribution is shown in part (c) of Fig. 1.  
Here, HIJING/B provides
a factor of 3 more valence $\Lambda$ at mid-rapidity consistent 
with the expected strangeness
enhancement associated with the baryon junction exchange.  
However the very large value of the valence hyperon yield
in $PbPb$ cannot be reproduced. Multiple final state interactions
or ``rope'' effects
must be included to understand those yields.
Increasing the effective string tension by a factor of two
makes little difference on the valence non-strange baryon yields,
but can account for the anomalous hyperon yield (dotted curves).

Finally in Fig 1d, the  predictions of this model 
for the valence proton rapidity distribution in $Au+Au$ collisions at 
RHIC energies ($\sqrt{s}$ = 200 GeV) are shown.  
HIJING/B predicts nearly twice the initial baryon density, $\rho_B(1\;{\rm fm})
\approx 2 \rho_0$, at mid-rapidity than HIJING.

In conclusion, a  novel  baryon junction stopping mechanism is implemented
in HIJING/B to reproduce baryon transport in $pp$ collisions.
This is sufficient to account
for the midrapidity
valence baryon data in $pA$ and $AA$. 
The model naturally
leads to
large enhancement of the $p_\perp$ slope
as well as of the yield of hyperons.
However, the preliminary  $\Lambda-\bar{\Lambda}$
 yield in $PbPb$ is still underestimated without multiple final state
 interactions. 


{\small Acknowledgements: We thank  Dima Kharzeev for  
stimulating discussions.
This work was supported in part by the Director, Office of Energy
Research, Division of Nuclear Physics of the Office of High Energy and Nuclear
Physics of the U.S. Department of Energy under Contract No. DE-FG02-93ER40764.}

\end{document}